\documentclass[twocolumn, twocolappendix]{aastex631}

\usepackage{subfigure}
\usepackage{graphicx}
\usepackage{multirow}

\begin{document}

\title{
How to feed your dragon: the role of turbulence in AGN activity 
}

\author[0000-0001-7522-1324]{Olga Borodina}
\affiliation{Center for Astrophysics $\vert$ Harvard \& Smithsonian, 60 Garden St, Cambridge, MA 02138, USA}

\author[0000-0002-8573-2993]{Jake S. Bennett}
\affiliation{Center for Astrophysics $\vert$ Harvard \& Smithsonian, 60 Garden St, Cambridge, MA 02138, USA}

\author[0000-0001-7899-7195]{Yueying Ni}
\affiliation{Michigan Institute for Data and AI in Society, University of Michigan, Ann Arbor, MI, 48109, USA}

\author[0000-0001-6260-9709]{Rainer Weinberger}
\affiliation{Leibniz-Institut fur Astrophysik Potsdam (AIP), An der Sternwarte 16, 14482 Potsdam, Germany}

\author[0000-0002-3301-3321]{Michaela Hirschmann}
\affiliation{Institute of Physics, Laboratory for galaxy evolution and spectral modelling, EPFL, Observatory of Sauverny, Chemin Pegasi 51, 1290 Versoix, Switzerland}

\author[0000-0003-4597-6739]{Boon Kiat Oh}
\affiliation{School of Physics, Korea Institute for Advanced Study, 85 Hoegiro, Dongdaemun-gu, Seoul 02455, Republic of Korea}

\author[0000-0001-6950-1629]{Lars Hernquist}
\affiliation{Center for Astrophysics $\vert$ Harvard \& Smithsonian, 60 Garden St, Cambridge, MA 02138, USA}


\begin{abstract}

Active galactic nuclei (AGN) are thought to be important drivers of galaxy evolution. However, the way AGN jets couple their energy to the interstellar medium (ISM) remains poorly understood. Recent observations have revealed $\sim$kpc-scale jets in galaxies of different morphological types.
Here, we use idealized turbulent box simulations with accreting black holes to investigate how jet–ISM interactions and feedback shape the evolution of jets.
We show that a turbulent ISM is a key ingredient in both jet launching and jet propagation. 
Without turbulence, jet feedback initially shuts off the cold gas supply to the black hole and the system enters a steady state with jet power $\leq 10^{39}\,\mathrm{erg\,s^{-1}}$ for a $10^{6}\,\mathrm{M}_\odot$ black hole. 
In a turbulent medium, jets fill voids with hot diffuse material which the turbulence then carries away, replenishing the black hole's fuel supply.
However, above a critical jet efficiency factor, the jet begins to significantly deplete its fuel supply and enters a self-limiting accretion regime. This critical factor depends on the ISM Mach number, with lower turbulence yielding a lower threshold.
Moreover, while the jet power depends on the ambient gas density in the vicinity of the black hole, we find that, at fixed turbulent structure, the morphology and extent of jet propagation are largely insensitive to the average background density; turbulence in the ISM dominates over mean density in setting how far jets propagate, although denser media experience higher total energy injection and stronger heating.

\end{abstract}



\section{Introduction} \label{sec:intro}

Cosmological simulations have shown that supermassive black holes (SMBHs) are key drivers of galaxy evolution \citep{Sijacki2007, DiMatteo2008, Booth2009, Sijacki2015}. In high-mass galaxies, the energy output from stellar feedback alone is insufficient to quench star formation \citep{Dubois2008}. 
Active galactic nuclei (AGN) provide an additional source of energy capable of heating the surrounding medium and driving large-scale outflows. These outflows can take the form of disk-driven winds \citep[e.g.][]{Silk1998} and relativistic jets \citep[e.g.][]{Blandford1982, Begelman1984}, where the former are often associated with radiatively efficient AGN and the latter with radiatively inefficient ones.

AGN jets can be launched by spinning SMBHs, powered by the extraction of black hole spin energy through magnetic fields that are supplied by the accreting gas \citep{Blandford1977}. Assuming the accretion disk is magnetically arrested, accretion onto the SMBH sets the energy available for this feedback. Yet the feeding of the black hole itself can be strongly influenced both by the surrounding gas conditions and by the feedback itself.

The interplay between the ISM and AGN is complex, and the conditions of the ambient ISM strongly influence how gas is transported to the SMBH. In particular, the level of turbulence on ~\,kpc scales can by itself regulate the black hole's feeding \citep{Barbani2026a, Barbani2026b, Cammelli2026a, Cammelli2026b}. Supernova (SN) feedback, one of the sources of this turbulence, can affect BH growth in different ways \citep[and references therein]{Alexander2012}. On the one hand, stellar feedback can limit accretion onto the black hole by heating and expelling gas \citep{Dubois2015, Habouzit2017, AnglesAlcazar2017, Petersson2025}, and dense star-forming regions can divert gas from the nucleus \citep{GonzalezDelgado2001, Trebitsch2018}. On the other hand, stellar feedback can drive winds that funnel gas toward the BH and enhance the accretion rate \citep{Hopkins2012, Partmann2025}, and stellar ionizing radiation can further boost the inflow \citep{Petersson2025}. In the maintenance mode as well, different strengths of turbulence in the medium on ~kpc scale can affect gas transport to the SMBH \citep{Cammelli2026a, Cammelli2026b}.

Conversely, SMBHs can significantly affect the ISM. AGN feedback can heat, expel, or redistribute gas, thereby cutting off the supply of material for both star formation and further accretion onto the BH. As a result, jet outflows can regulate their own power \citep{Wyithe2003, KungYi2023}, leading to a self-regulated duty cycle. 
Intermittent jet activity has been investigated in several studies and has been attributed to, for instance, magnetic instabilities and jet-induced turbulence \citep{Ressler2021, Lalakos2024}.

Observations provide strong evidence for interactions between the ISM and AGN \citep{Hardcastle2012, GarciaBurillo2014, Kharb2019, Kharb2021, Fabbiano2022}. For example, low-power jets, such as FR0 sources \citep{Baldi2016, Baldi2018, Baldi2023}, are stalled on kiloparsec scales for millions of years, allowing ample time for jet energy to influence ISM properties. When resolved, these jets exhibit complex morphologies with asymmetric structures \citep{Giovannini2023} which can be linked to the jet-ISM interaction \citep{Sutherland2007, Mukherjee2016, Mukherjee2018, Tanner2022, Borodina2025}; also see \citet{Mukherjee2025} for a recent review. Furthermore, \citet{Murthy2022} showed that even low-luminosity AGN can expel most of the cold gas out to $\sim$500\,pc, potentially starving the black hole.

General-relativistic magnetohydrodynamic (GRMHD) simulations can model the launching of jets and outflows from first principles \citep[e.g.][]{Tchekhovskoy2011, Narayan2012, Liska2018, Liska2020}.
Using these simulations, \citet{Narayan2012, Yuan2014} showed that hot accretion onto a non-spinning black hole can still launch collimated, quasi-relativistic jets via the disk-driven Blandford–Payne mechanism \citep{Blandford1982} together with wide-angle winds. These winds reach much lower velocities than the jet but carry a higher mass and energy flux \citep{Yuan2015}. For spinning black holes the outflow can also include both relativistic jets and broader winds \citep{McKinney2014, Narayan2022}.
Beyond GRMHD, hydrodynamical simulations that prescribe non-collimated outflows on larger scales have also been used to study how wide-angle jets and winds couple to the ambient medium \citep[e.g.][]{Gourgouliatos2018, Ward2024, Dong2025}.


SMBH feeding and feedback are in constant interplay, which is why they must be studied jointly \citep{Gaspari2020}.
However, connecting AGN jet launching to its galaxy-scale feedback effects is challenging because no single method spans the full range of scales. Conventional GRMHD simulations cannot reach the Bondi radius for realistic gas temperatures \citep{Narayan2022}, and even multi-zone GRMHD models \citep{Cho2024, Guo2025} do not extend to kiloparsec scales.
Galaxy-scale simulations, conversely, capture the overall impact of AGN feedback but rarely resolve Bondi-scale accretion without becoming prohibitively expensive. Targeted and super-Lagrangian refinement is beginning to bridge this gap, resolving the Bondi radius in idealized simulations of galaxies \citep{Curtis2015, Koudmani2019}, in galaxy clusters and groups \citep{Bourne2017, Barbani2026a, Barbani2026b, Cammelli2026a, Cammelli2026b}, and in some cosmological runs \citep{AnglesAlcazar2021}. To date, however, these either omit AGN feedback or lack a well-resolved, multiphase turbulent ISM, which remains prohibitively expensive in full cosmological volumes.
A complementary route is to model the unresolved physics: \citet{Su2025} applied the \citet{Cho2024} multi-zone GRMHD solution as a sub-grid accretion-and-feedback prescription in galaxy-scale simulations, but found that this alone does not reproduce realistic black hole growth.

A handful of simulations have modeled jets at enhanced resolution relative to the surrounding medium, while coupling them to self-consistent accretion prescriptions \citep{Talbot2021, Talbot2022}. In these works, a sub-grid accretion disc model is used to launch the jet and to follow its interaction with a resolved circumnuclear disc. \citet{Talbot2021} show that the jet regulates the gas supply to the black hole, leading to an evolution that differs from fixed–jet-power simulations. However, the surrounding medium in these simulations does not represent a well-resolved, multiphase turbulent ISM which can affect both accretion and jet propagation.

\citet{Borodina2025} showed that turbulent gas drives the jet propagation and jet morphology  unless jet power exceeds the dynamic pressure of the gas. Also, \citet{Borodina2025} found that for the jet power of $\leq 10^{40}$\,erg\,s$^{-1}$ the jet can be stalled in the central kpc region. However, this initial study only investigated constant jet powers, their propagation, and their impact on the ISM, and did not consider the back-reaction on accretion.

In this work, we aim to bridge this gap by performing idealized simulations that resolve the accretion region surrounding the black hole, within which the Bondi radius is contained, and self-consistently follow both accretion and feedback for BHs. The simulation setup is described in Section~\ref{sec:methods}. In Section~\ref{sec:results}, we present the results of our analysis, distinguishing the effects of jet and ISM properties. Section~\ref{sec:discussion} discusses how our findings relate to observational data and previous simulations. Finally, we summarize our conclusions in Section~\ref{sec:conclusion}.

\section{Methods} \label{sec:methods}

We perform hydrodynamical simulations of AGN driven jets interacting with a turbulent interstellar medium environment with the \texttt{Arepo} code \citep{Springel2010, Pakmor2016}, which solves the Euler equations using a finite-volume method on a quasi-Lagrangian moving Voronoi mesh.

\subsection{Turbulent box} \label{sec:methods_turb}

We set up the initial turbulent box following the procedure described in \citet{Borodina2025}. The simulation box has a size of 2\,kpc and a resolution of $256^3$ cells, with periodic boundary conditions. To minimize unphysical effects associated with these boundaries, our analysis focuses on the central (1\,kpc)$^3$ region. We initialize the box with a uniform medium of a prescribed mean number density and an initial temperature of $10^4$\,K, and drive turbulence by injecting energy corresponding to a specified Mach number \citep{Bauer2012, Mocz2018}. At the adopted mean number density of $20\ \mathrm{cm}^{-3}$, this resolution corresponds
to a target gas mass of $\sim 150\ M_\odot$ per cell; varying the initial mean density
therefore implies different mass resolutions across simulations. We tested the impact of this varying mass resolution in our previous work \cite[see][]{Borodina2025} and found no significant effect on the main diagnostics we consider, so we do not expect it to alter the conclusions presented here.

We include radiative cooling \citep{Katz1996}, assuming primordial abundances and imposing a temperature floor of $10^{4}$\,K. As the driven turbulence develops, the gas evolves into a medium with a broad range of densities and temperatures; we show the corresponding mass-weighted distributions for different Mach numbers in Appendix~\ref{app:hists}.

In this work, we do not include physics such as gravity, magnetic fields, or relativistic effects. For systems like FR0 jets on the scales of interest, general relativistic effects can be neglected \citep{Giovannini2023}, although we plan to incorporate special relativistic effects in future work.
The role of magnetic fields remains uncertain: some studies suggest that strong magnetic fields can distort jet lobes, while others argue that they suppress mixing between jet material and the ambient medium \citep[see][and references therein]{Bourne2023}. Moreover, magnetic field strengths in galactic centers are poorly constrained, making their inclusion highly uncertain.
We therefore focus on the hydrodynamic interaction between jets and the ISM, deferring additional physical processes, such as magnetic fields and gravity, to future studies.

The strength of turbulence is controlled by the specific energy injection rate, $\epsilon$, which we calculate using the following equation \citep{Mocz2018}:

\begin{equation}
    \epsilon = \alpha \cdot \frac{c_s^3}{ 2^3 \cdot l},
    \label{eq:energy}
\end{equation}
where $c_s$ is the sound speed, and $l$ is the turbulence driving scale, which is set to $l = 1$~kpc. The parameter $\alpha$ is the energy-loading factor and is varied to achieve the desired Mach number.

The jet is launched when the box reaches the steady state which depends on the largest eddy turnover time (as described in \citealt{Borodina2025}). The estimated times and energy loading factors are described in Table~\ref{tab:turbbox}.

\begin{table}[h!]
    \centering
    \caption{Parameters of the turbulent boxes for the different $10^4$\,K gas target Mach numbers.}
    \begin{tabular}{ccc}
    \hline
       Mach number  &  $\alpha$ & Time of jet launch, Myr\\
       \hline
        4 &  1 & 15\\
        2 & 0.3 & 25\\
        1 & 0.01 & 50\\
        0.5 & 0.001 & 100\\
        \hline
    \end{tabular}

    \label{tab:turbbox}
\end{table}

\begin{table*}[ht!]
\centering
\caption{Summary of simulation parameters for the different experiments. Runs marked with an asterisk ($^*$) were repeated with several turbulent realizations.}
\begin{tabular}{lcccc}
\hline
Experiment & Efficiency Factor $\varepsilon$ & Opening Angle $\theta$ ($^\circ$) & Mean ISM Density $n$ (cm$^{-3}$) & Mach number $\mathfrak{M}$ \\
\hline

\multirow{10}{*}{\textit{Varying $\varepsilon$}} 
  & $10^{-4}$ & 0 & 20 & 4 \\
  & $10^{-3}$ & 0 & 20 & 4 \\
  & $10^{-2}$ & 0 & 20 & 4$^*$\\
  & $10^{-1}$ & 0 & 20 & 4 \\
  & $10^{0}$  & 0 & 20 & 4 \\
\cline{2-5}
  & $10^{-6}$ & 0 & 20 & 1 \\
  & $10^{-5}$ & 0 & 20 & 1 \\
  & $10^{-4}$ & 0 & 20 & 1 \\
  & $10^{-3}$ & 0 & 20 & 1 \\
  & $10^{-2}$ & 0 & 20 & 1$^*$\\
\hline

\multirow{3}{*}{\textit{Varying $n$}} 
  & $10^{-2}$ & 0 & 100  & 4 \\
  & $10^{-2}$ & 0 & 20 & 4$^*$ \\
  & $10^{-2}$ & 0 & 1  & 4 \\
\hline

\multirow{5}{*}{\textit{Varying $\mathfrak{M}$}} 
  & $10^{-2}$ & 0 & 20 & 4$^*$ \\
  & $10^{-2}$ & 0 & 20 & 2 \\
  & $10^{-2}$ & 0 & 20 & 1 \\
  & $10^{-2}$ & 0 & 20 & 0.5 \\
  & $10^{-2}$ & 0 & 20 & 0 \\
\hline

\multirow{4}{*}{\textit{Varying $\theta$}} 
  & $10^{-2}$ & 0   & 20 & 4$^*$ \\
  & $10^{-2}$ & 180 & 20 & 4 \\
  & $10^{-2}$ & 0   & 20 & 1 \\
  & $10^{-2}$ & 180 & 20 & 1 \\
\hline

\end{tabular}
\label{tab:sims}
\end{table*}

\subsection{Accretion-regulated jet launching} \label{sec:methods_jet}

The jet is launched from the central region of the simulation box with a fixed injection radius of 90\,pc, following the prescription of \citet{Weinberger2017b, Weinberger2023}. 
To study the role of the angular distribution of the momentum injection in the jet launching region, we consider two models in this paper: the first is a collimated jet, in which the momentum imparted to each cell is aligned with the $x$-axis. The second is a spherically symmetric wind, where each particle is accelerated radially outward from the black hole. 

The jet injection region consists of two zones. The inner zone is a central sphere of radius 30\,pc, from which the jet is launched. Part of the jet energy is first deposited thermally, to ensure pressure equilibrium between the jet injection region and the surrounding material. Then, the rest of the energy (the dominant part) is deposited kinetically. We set the jet density within the inner sphere as $10^{-26}$\,g\,cm$^{-3}$. The outer zone, referred to as the accretion zone, is a spherical shell extending from 30 to 90\,pc in radius. This accretion zone is used to estimate the average ambient gas pressure, density, and sound speed. 
We then estimate the accretion rate $\dot{M}$ using the Bondi-Hoyle-Lyttleton (BHL) formula \citep{Bondi1952}:

\begin{equation}
     {\dot {M}_\mathrm{BHL}}= \frac{4 \pi \langle \rho \rangle G^{2}M^{2}_\mathrm{BH}}{(\langle c_{s} \rangle ^{2} + \langle v_\mathrm{gas} \rangle ^{2})^{3/2}},
     \label{eq:Bondi}
\end{equation}
where $\langle \rho \rangle$ is the kernel-weighted gas density, $M_\mathrm{BH}$ is the black hole mass, and $\langle c_{s} \rangle$ and $\langle v_\mathrm{gas} \rangle$ are the kernel-weighted sound speed and gas velocity (relative to the black hole) in the accretion region. 

We note that these choices have several important consequences. First, the cells containing jet material are also included in the averaging. As a result, once the jet is energetic enough to break out of the injection region into the outer shell of the accretion zone, the hot, low-density jet gas lowers the inferred accretion rate. Because jet energy coupling operates close to our resolution scale, its strength may be overestimated. We nonetheless retain jet material in the averaging: excluding it would implicitly assume that the jet does not interact with the ambient gas at these scales, which would be a less physical assumption. Second, a further limitation is inherent to the BHL formalism itself rather than to our implementation of accretion. For a symmetric outflow, e.g., a spherical wind or a bipolar jet, the outward-moving gas contributes with opposite signs in different directions and largely cancels in this vector average. The formula therefore cannot capture the suppression of accretion by a divergent velocity field, even though outward-moving gas physically reduces the inflow. A related effect is that turbulent motions on scales below our resolution limit are not captured, which can likewise bias the computed accretion rate \citep{Krumholz2004, Krumholz2006}. As a result, we may overestimate accretion in the outflow-dominated regime.

Once $\dot{M}$ is determined, the corresponding gas mass is drained from all cells within the accretion zone, weighting the removal toward cells that are closer to the black hole and that have larger volumes. The black hole sink particle carries a subgrid mass that grows at the rate $\dot{M}$, while gas is physically removed from the grid only once this subgrid mass has caught up with the black hole's dynamical (gravitational) mass \citep{Springel2005}.

Although cosmological simulations commonly adopt the Bondi and Bondi-Hoyle-Lyttleton as a simplified sub-grid estimate of black hole accretion, GRMHD calculations suggest that the actual accretion rate is typically lower and can be approximated as a fraction of the Bondi rate \citep{Lalakos2022, Cho2025, Guo2025}. For example, \citet{Cho2025} showed that the efficiency factor is approximately $\varepsilon \approx 10^{-5}$ for a black hole of mass $10^6\,M_\odot$. However, this relation is not valid for gas with temperatures below $10^5$\,K. Consequently, the fraction of the Bondi accretion rate that may actually be accreted onto the black hole remains uncertain.

We therefore introduce an efficiency factor, $\varepsilon_t(t)$, to modulate the Bondi accretion rate. On the timescale of our simulations (i.e., neglecting black hole mass growth via accretion), this factor is degenerate with both the radiative efficiency and the jet feedback efficiency. Therefore, for the purposes of this work we combine all of these into a single efficiency factor, $\varepsilon$, defined as:

\begin{equation}
    L_\mathrm{jet}  = \varepsilon_t (t) \dot{M}_\mathrm{BHL} c^2,
    \label{eq:jetpower}
\end{equation}
where $L_\mathrm{jet}$ is the jet power, $c$ is the speed of light, and $\varepsilon_t(t)$ is an efficiency factor that increases over time to reach the target value of $\varepsilon$, as described below.

The initial gas density in the center of the simulation box, where the jet injection region is located, can reach high values and cause an unrealistically powerful burst of feedback at the beginning of the jet launching. We therefore gradually activate the jet to clear out the surrounding gas in a controlled manner. To do so, we introduce a sigmoid activation function that increases the efficiency from zero up to the target value $\varepsilon$ in 0.5\,Myr. Moreover, we apply an additional refinement criterion on the jet material, so that every cell containing jet tracer above a threshold value of $10^{-3}$ is refined to our maximum resolution of $\sim$\,9 pc.


We set the initial black hole mass to $M_\mathrm{BH} = 10^6\,M_\odot$, which ensures that the Bondi radius for $T = 10^4$\,K gas is resolved and lies within our 30–90\,pc accretion region, so that the accretion rate is evaluated over gas that approximately encompasses the Bondi sphere. As gas in the vicinity of the black hole is heated, the Bondi radius shrinks down to sub-pc scale, keeping the Bondi sphere within the accretion region at all times, although it is no longer resolved.

\section{Results} \label{sec:results}

We performed a suite of simulations varying four free parameters. Two of them describe the ISM properties: the mean number density, $n$, and the Mach number, $\mathfrak{M}$, as introduced in Section~\ref{sec:methods_turb}. The other two parameters characterize the jet properties: the jet efficiency, $\varepsilon$, and the jet opening angle, $\theta$, introduced in Section~\ref{sec:methods_jet}. Table~\ref{tab:sims} summarizes the simulation suite. In this section, we describe how variations in each parameter affect the system’s evolution.

\subsection{Efficiency factor significance} \label{sec:results_epsilon}

First, we vary the efficiency factor $\varepsilon$ from $10^{-6}$ to $10^{0}$. This parameter is not confidently constrained for multiphase gas from first principles. Therefore, we aim to understand how the simulation behaves under different choices of this efficiency factor.

As long as the jet does not significantly alter the gas near the black hole, its power increases linearly with the efficiency factor defined in Eq.~\ref{eq:jetpower}.  However, when the jet becomes sufficiently powerful, it alters the inflow and its power is correspondingly reduced, a behavior we refer to as \textit{self-limiting accretion}.

\begin{figure*}[ht!]
    \centering
    \includegraphics[width=\textwidth]{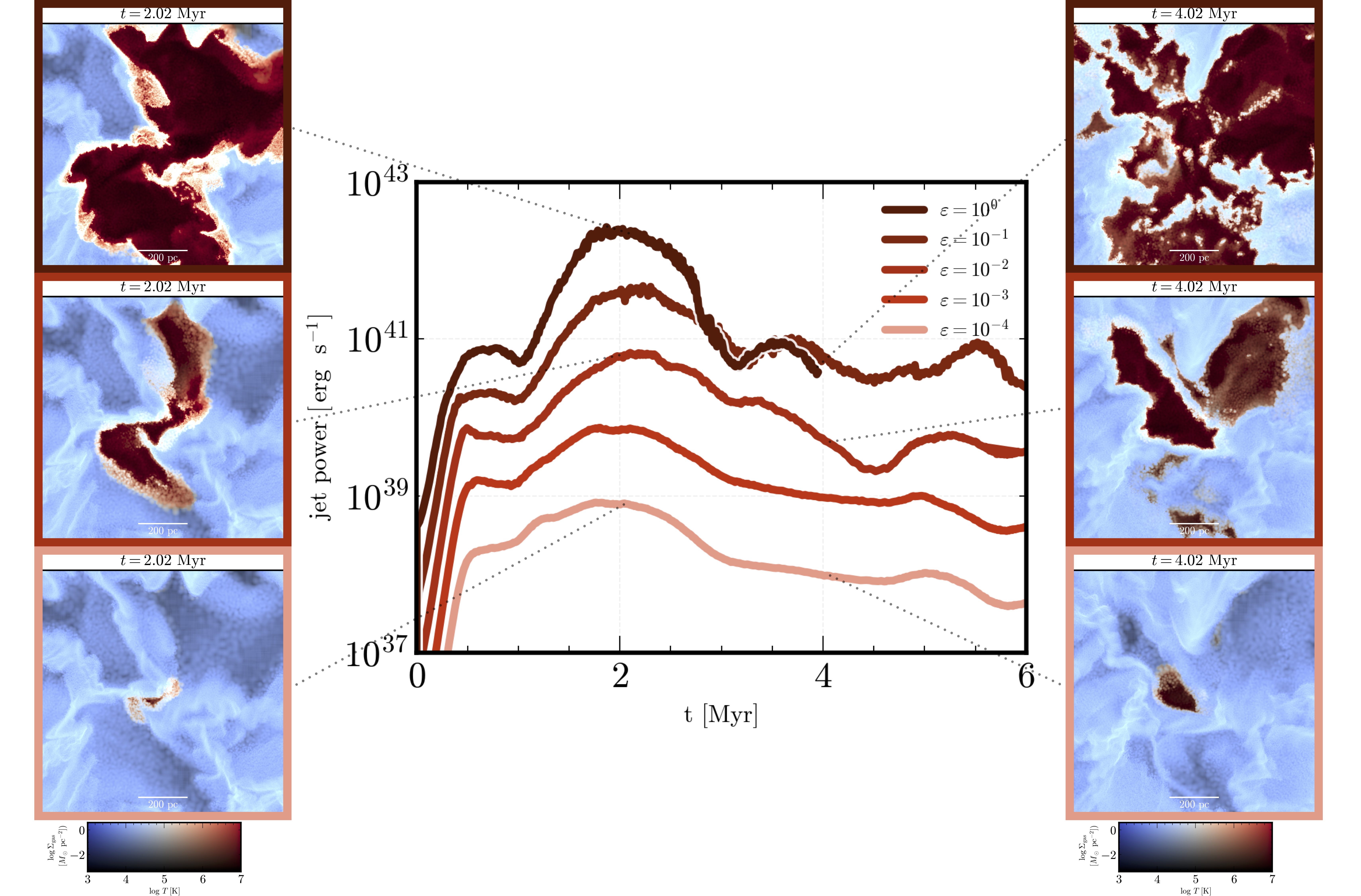}
    \caption{\textit{Central panel}: Jet power evolution for models with different jet efficiency factor values.
    \textit{Left panels}: Density-temperature projection maps of simulations with $\varepsilon=[10^0, 10^{-2}, 10^{-4}]$ at 2 Myr after the start of jet launching.
    \textit{Right panels}: Density-temperature projection maps of simulations with $\varepsilon=[10^0, 10^{-2}, 10^{-4}]$ at 4 Myr after the start of jet launching.    
    All the simulations have the same ISM conditions: $n = 20$ cm$^{-3}$, $\mathfrak{M} = 4$. The jet opening angle is set to $0^\circ$.}
    \label{fig:epsilon}
\end{figure*}

In Fig.~\ref{fig:epsilon}, we show the evolution of jet power for models with different efficiency factors, together with density–temperature projection maps at selected snapshots. At lower efficiencies, $\varepsilon<10^{-1}$, there is no sign of self-limiting accretion; in this regime the jet power increases approximately linearly with $\varepsilon$. Moreover, the time variations of the jet power are synchronous for efficiency factor between $\varepsilon = 10^{-4}$ and $\varepsilon = 10^{-3}$, indicating that the variability in the jet power is driven by \textit{the turbulent medium itself} rather than by jet feedback.

However, at sufficiently high efficiency, $\varepsilon > 10^{-1}$, the jet reaches jet power of $10^{42}$\,erg\,s$^{-1}$ and begins to clear out the gas that would otherwise fuel the AGN, causing the jet power to drop. We cannot determine whether this reduction is sustained as the more powerful jet eventually reaches the simulation boundary and the run must be stopped (indicated by the early cut of lines with $\varepsilon= 10^{0}$). We demonstrate this convergence explicitly at lower Mach number in Appendix~\ref{app:self-lim-M}. This result suggests that at high efficiency the jet may be capable of limiting accretion, although in some cases it does not remain confined within the central 1\,kpc.

In the left and right panels of Figure~\ref{fig:epsilon}, we show projection maps capturing the jet state at two different times. For $\varepsilon = 10^{-4}$, the jet remains confined within the central 1\,kpc and only weakly disrupts the surrounding gas. Higher jet efficiencies allow the jet to propagate beyond the central 1\,kpc. In particular, for $\varepsilon = 10^{0}$, the jet strongly disrupts the ISM and effectively cuts off its own fuel supply, leading to self-limiting accretion.

\subsection{Dependence on the ISM density}

Galaxies that host jets exhibit a wide range of central ISM densities \citep{Werner2014, Kakkad2018}, which can significantly affect jet evolution. In Seyfert galaxies at redshift $z < 0.2$, the observed electron number density is $\sim100\,\mathrm{cm^{-3}}$ on kiloparsec scales \citep{Kakkad2018}. In contrast, elliptical galaxies have lower ISM gas fractions \citep{BinneyJames2008GD}, with central electron densities down to $\sim0.07\,\mathrm{cm^{-3}}$ \citep{Werner2014}. Motivated by this range, in this work we vary the mean ISM number density among 1, 20, and 100\,$\mathrm{cm^{-3}}$ to investigate its impact on jet power and propagation.

\begin{figure}[ht!]
    \centering
    \includegraphics[width=0.94\linewidth]{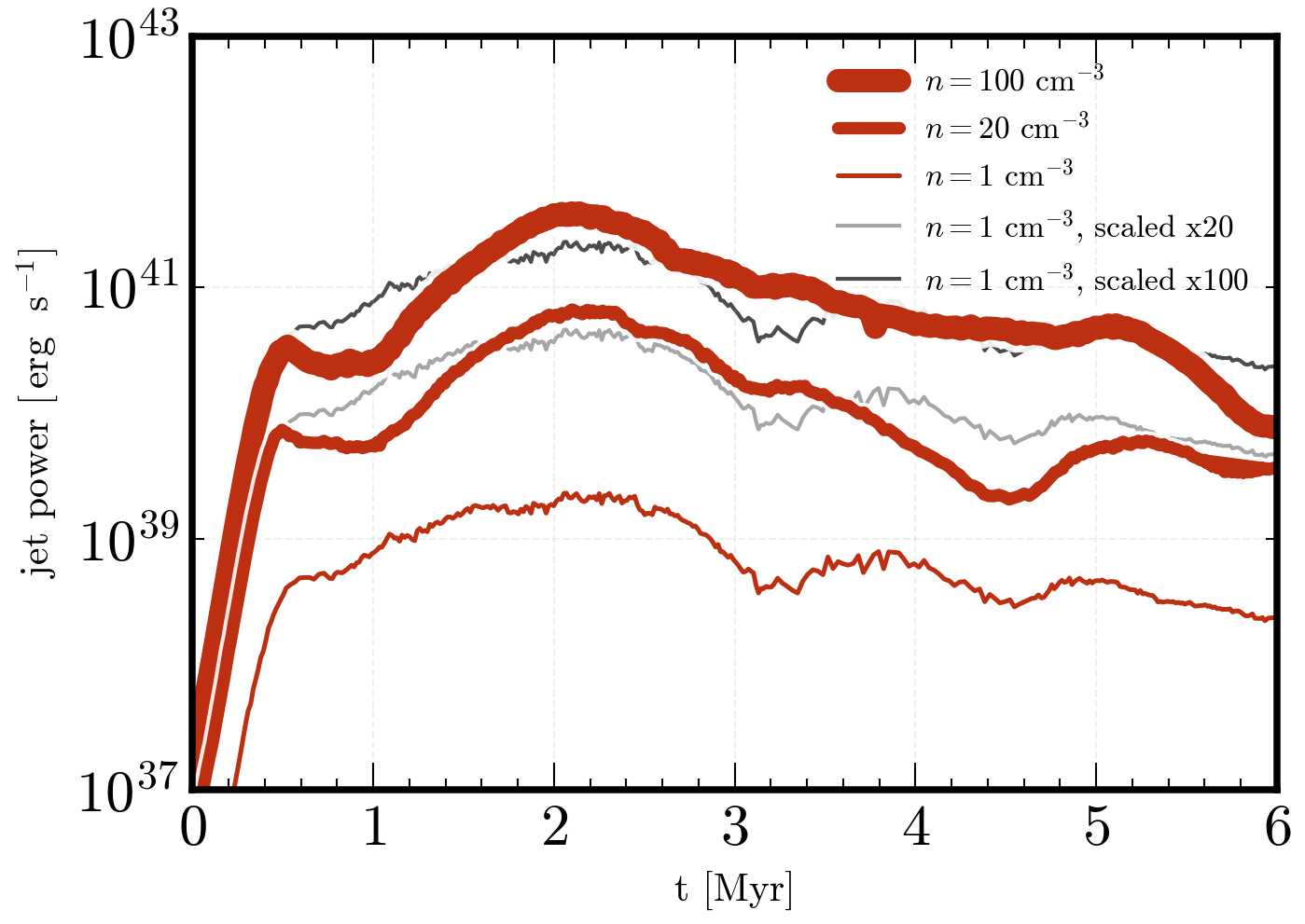}
    \caption{Jet power evolution for models with different mean ISM number densities. Jets have an efficiency factor $\varepsilon=10^{-2}$, and are launched into a medium with Mach number $\mathfrak{M} = 4$. The jet opening angle is set to $0^\circ$. The red lines of increasing thickness show the actual simulations at $n = 1,\,20,\,100\,\mathrm{cm^{-3}}$; the gray and black lines show the $n = 1\,\mathrm{cm^{-3}}$ run scaled by 20 and 100, respectively.}
    \label{fig:density}
\end{figure}

In Fig.~\ref{fig:density}, we show the evolution of jet power for simulations with different mean ISM number densities. When the jet is launched in a lower-density medium, its power is correspondingly lower. The difference in jet power between the higher- and lower-density environments matches the ratio of their mean number densities, as indicated by the similarity between the medium-thick red and thin gray lines as well as the thickest red and black lines. This behavior arises from our choice of accretion prescription: the Bondi-Hoyle-Lyttleton accretion rate scales linearly with the mean density, because the sound speed for the cells in the jet injection region is the same across these simulations and the black hole mass does not change significantly over the simulation time. Moreover, because the efficiency factor is not high enough to reach self-limiting accretion, jet power depends strongly on the state of the ambient medium.

\begin{figure*}[ht!]
    \centering
    \includegraphics[width=0.9\textwidth]{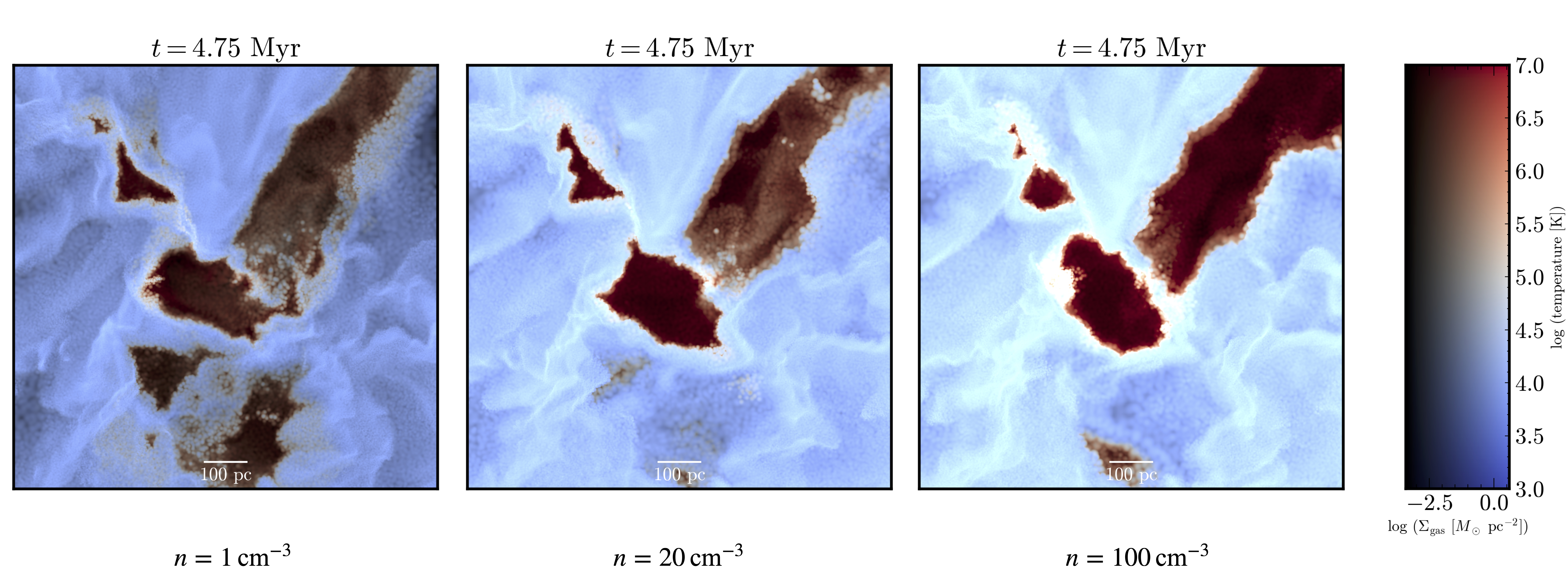}
    \caption{Projection maps (100\,pc in width) of simulations with jets launched in a turbulent ISM with different mean number densities. The left panel shows $n = 1\,\mathrm{cm^{-3}}$, the middle panel $n = 20\,\mathrm{cm^{-3}}$, and the right panel $n = 100\,\mathrm{cm^{-3}}$. The jets have an efficiency factor of $\varepsilon = 10^{-2}$ and are launched into a medium with Mach number $\mathfrak{M} = 4$. The jet opening angle is set to $\theta = 0^\circ$.}
    \label{fig:density_propagation}
\end{figure*}

Despite a difference of up to two orders of magnitude in jet power between these runs, we find that propagation is primarily determined by the turbulence. This contrasts with our earlier work with fixed-power jets \citep{Borodina2025}, where denser gas more effectively stalled the jet. Here, the Bondi-like accretion prescription causes the jet power to scale with the mean density, compensating for the increased confining pressure and rendering propagation roughly density-independent.
In Fig.~\ref{fig:density_propagation}, we present projection maps at the same time for three simulations with low, medium, and high mean number densities. The turbulent structure is similar in all three cases, and the jet primarily fills pre-existing cavities, so its subsequent propagation is governed by the turbulent ISM rather than by the mean density. Nevertheless, the higher jet power in the denser medium leads to a larger total injected energy and more efficient heating of the gas within the cavities, even though the cavities do not expand to substantially larger radii over the times shown.

\subsection{Dependence on the ISM Mach number}

The structure of the ISM is a key property of the surrounding medium and can be characterized by the Mach number, which sets the density and velocity distribution of the gas as well as the eddy turnover time. We vary the Mach number as $\mathfrak{M} = 0, 0.5, 1, 2,\ \text{and}\ 4$, where $\mathfrak{M} = 0$ corresponds to a uniform, non-turbulent medium.

\begin{figure*}[ht!]
    \centering
    \includegraphics[width=\textwidth]{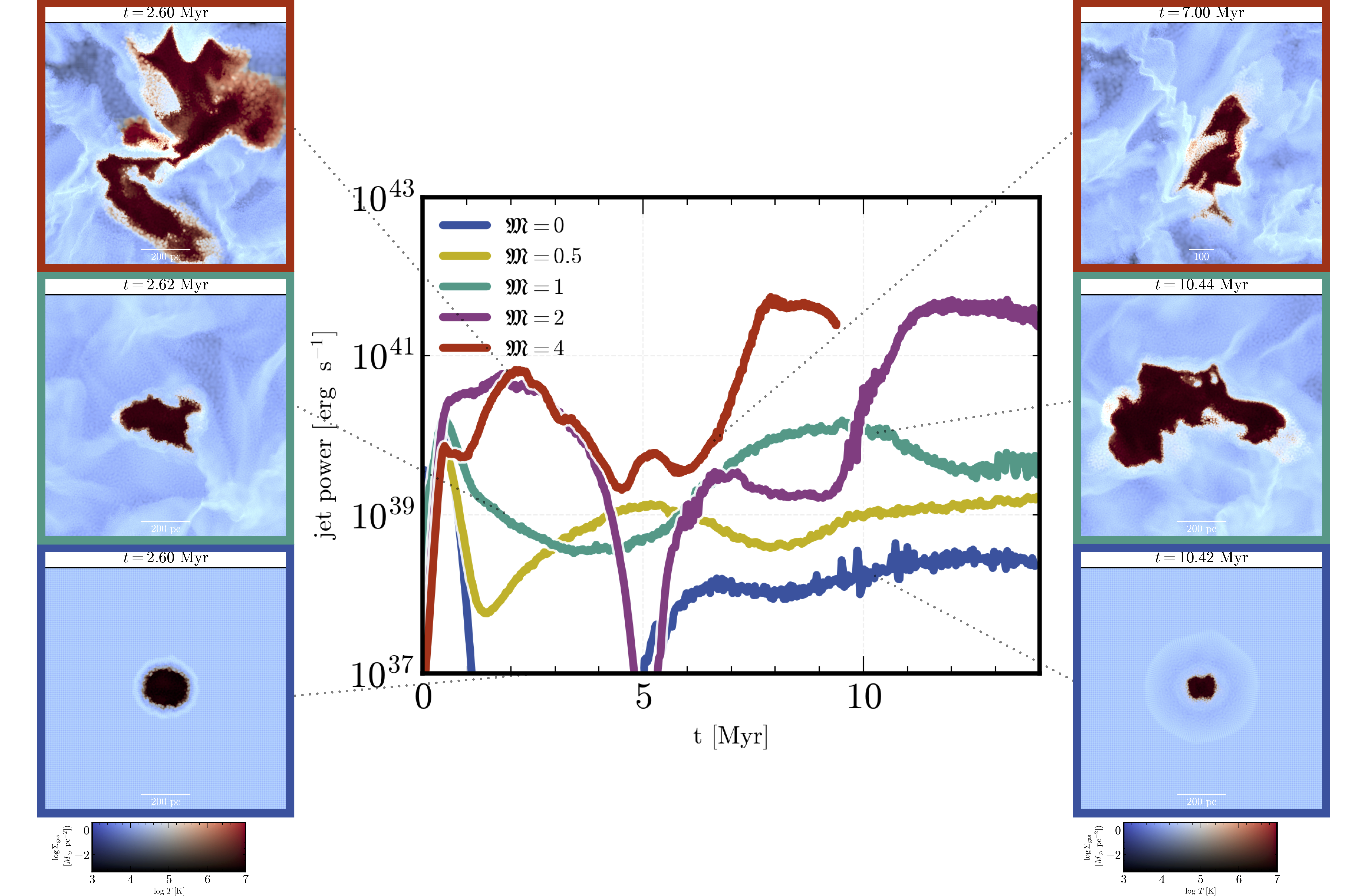}
    \caption{
    \textit{Central panel}: Jet power evolution for models with different ISM Mach numbers. 
    \textit{Left panels}: Density-temperature projection maps of simulations with $\mathfrak{M}=[0, 1, 4]$ at 2.6\,Myr after the start of the jet launching.
    \textit{Right panels}: Density-temperature projection maps of simulations with $\mathfrak{M}=[4]$ at 7\,Myr and $\mathfrak{M}=[0, 1]$ at 10.4\,Myr after the start of the jet launching.   
    All the simulations have the same ISM mean number density: $n = 20$ cm$^{-3}$, and the jet has an efficiency of $\varepsilon=10^{-2}$ and an opening angle of $0^\circ$.}
    \label{fig:mach}
\end{figure*}

In Fig.~\ref{fig:mach}, we show the evolution of jet power for jets launched into ISM environments with different Mach numbers. For the subsonic and transonic cases, $\mathfrak{M} = 0.5$ and $\mathfrak{M} = 1$, the jet can choke itself after the initial burst, although the power does not drop below $\approx 10^{38}$\,erg\,s$^{-1}$. After this initial decline, the jet launched in the $\mathfrak{M} = 1$ medium behaves similarly to the supersonic runs: the jet power is highly variable and can reach at least $10^{40}$\,erg\,s$^{-1}$. The jet in the subsonic medium, on the other hand, reaches a steady state after $\sim4$\,Myr, sharing characteristics with jet propagation in a uniform medium, which we describe in more detail below.

In the uniform medium ($\mathfrak{M} = 0$), the jet initially clears out the surrounding gas, driving the power down below $10^{37}$\,erg\,s$^{-1}$(corresponding to the Bondi rate in a hot, diffuse medium). In the absence of sustained jet driving, the hot bubble produced during this phase gradually cools and mixes with the ambient gas. After $\sim5$\,Myr, the cooled gas refills the injection region and triggers jet reactivation. At this stage, the dilute cocoon allows the jet to become more easily collimated (see the difference between the lower panels on the left and right in Fig.~\ref{fig:mach}). Gas inflow from lateral directions continues to feed the AGN, leading to a quasi-steady equilibrium regulated by the cooling rate.

We also note that the morphology of the jet in a uniform medium is characterized by a nearly symmetric cocoon, which is very different from jets in a non-uniform ISM. Even weak turbulence is sufficient to perturb the jet and produce asymmetric structures.

Supersonic turbulence enables the jet to reach higher power levels. For the $\mathfrak{M} = 4$ case, the jet power varies around $10^{40}–10^{42}$\,erg\,s$^{-1}$. For the lower Mach number, $\mathfrak{M} = 2$, the jet temporarily chokes itself at $\sim$5\,Myr, but turbulence still replenishes gas in the injection region. This trend persists across different turbulent seeds (Appendix~\ref{app:seeds}). 

\begin{figure}[ht!]
    \centering
    \includegraphics[width=\linewidth]{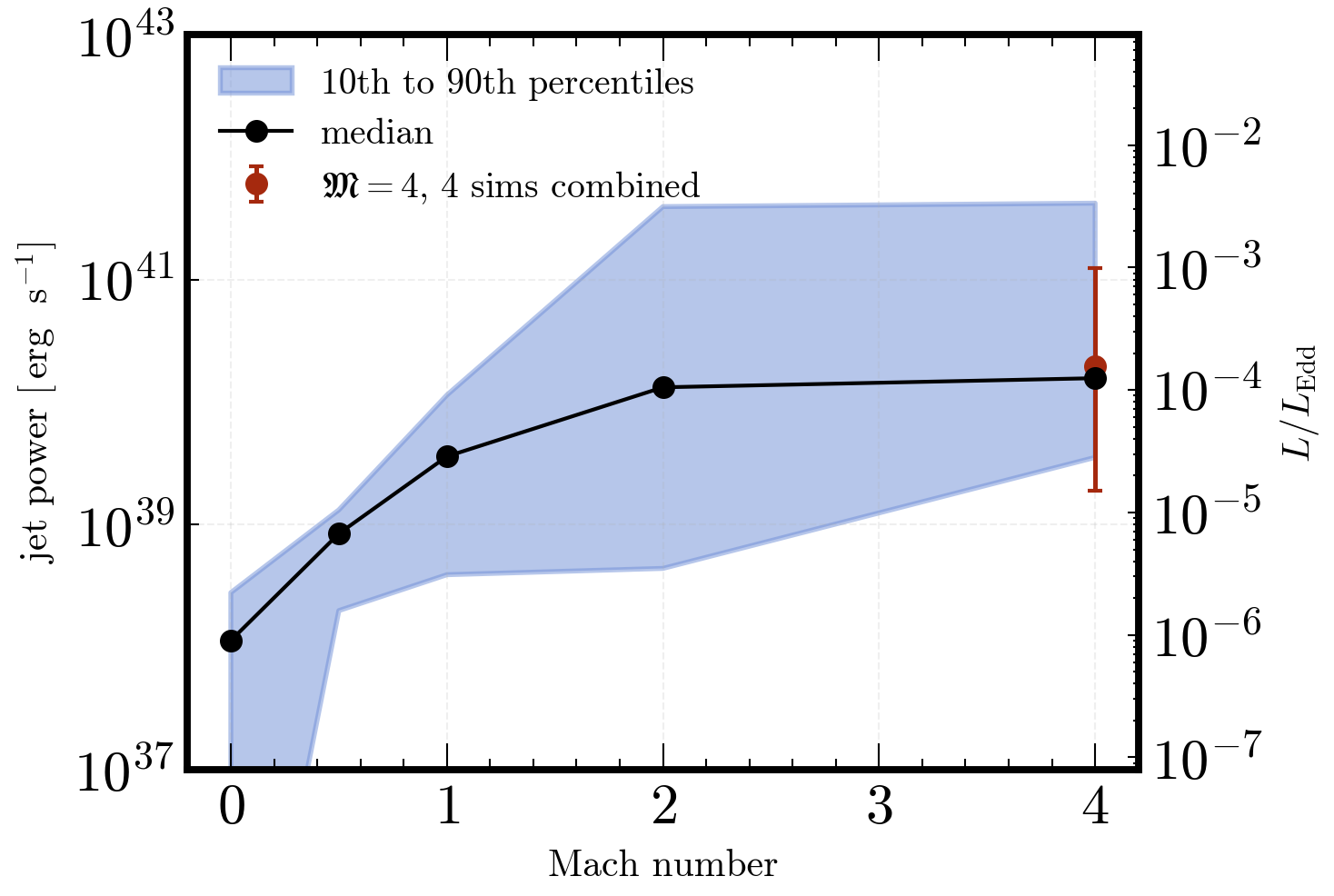}
    \caption{The 10th, 50th, and 90th percentiles of jet power for simulations with different Mach numbers. The first 1\,Myr is excluded from the averaging. The black line represents the median, and the blue shaded region highlights the scatter in jet power between the 10th and 90th percentiles. Red point and error bars represent 50th and 10th, 90th percentiles respectively for four simulations with the same parameters choices but different initial conditions. }
    \label{fig:machjetpower}
\end{figure}

In Fig.~\ref{fig:machjetpower}, we show the evolution of the median jet power for simulations with different Mach numbers. The median jet power increases as the ISM becomes more turbulent. In the uniform medium, the measured scatter in jet power is large because there is an extended period between 1 and 5\,Myr during which the jet is quenched; once the system reaches a steady state, the scatter becomes minimal. The scatter increases up to $\mathfrak{M} = 2$ because turbulence generates a structure with a broad density distribution: when a low-density cavity approaches the injection region, the jet power decreases, whereas it increases when a dense filament moves in. This duty cycle becomes less pronounced in the $\mathfrak{M} = 4$ ISM, where cavities and filaments move rapidly through the jet injection region, reducing the time the jet spends in either extreme accretion state. For $\mathfrak{M} = 4$, we also ran four additional simulations with different initial conditions and find that the scatter among all runs remains smaller than for the $\mathfrak{M} = 2$ case.

Although the $\mathfrak{M} = 0$ case has a non-zero jet power, the jet does not propagate beyond the bubble created by the initial outburst (see lower-right panel in Fig.~\ref{fig:mach}) over the $\sim15$\,Myr spanned by our simulations. In contrast, jets in turbulent media continue to propagate beyond 400\,pc. We note that in the projection map for the  $\mathfrak{M} = 4$ ISM at 7\,Myr (see upper-right panel in Fig.~\ref{fig:mach}), a jet bubble has already left the 1\,kpc region. As a result, the higher the turbulence, the more efficient jet propagation is, as previously shown by \citet{Borodina2025} for the fixed jet power.

\begin{figure*}[ht!]
    \centering
    \includegraphics[width=0.9\textwidth]{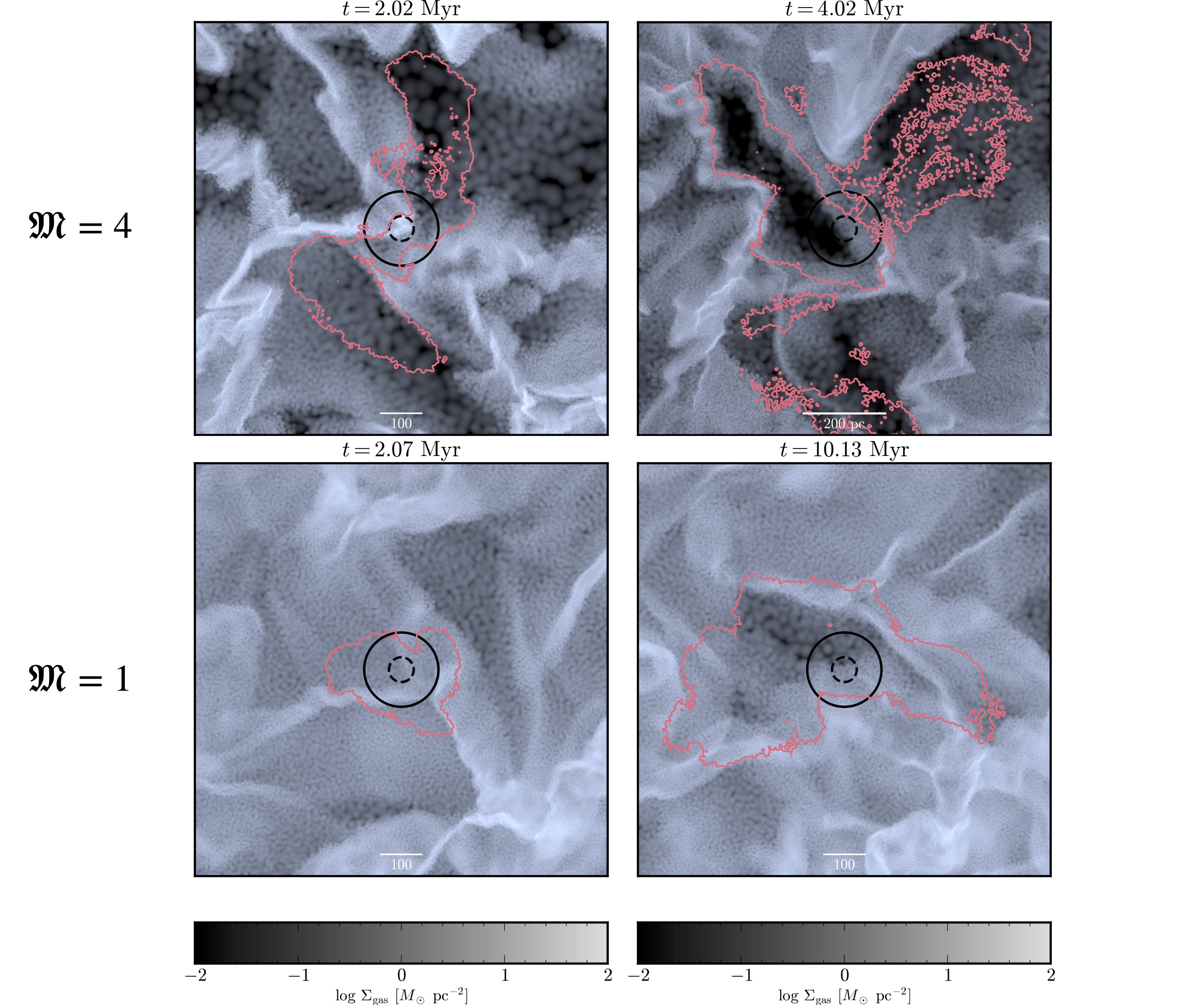}
    \caption{
    Projection maps of gas density in turbulent boxes (50\,pc in depth) without jets, with jet contours overlaid from simulations that include jet launching. Dashed and solid black lines represent location of jet injection and accretion zones, respectively.
\textit{Top panels:} Simulations with $\mathfrak{M} = 4$ at 2 and 4\,Myr after the start of jet launching.
\textit{Bottom panels:} Simulations with $\mathfrak{M} = 1$ at 2 and 4\,Myr after the start of jet launching.}
    \label{fig:machcontours}
\end{figure*}

In Fig.~\ref{fig:machcontours}, we show projection maps of the turbulent boxes alone, i.e., with the ISM unperturbed by the jets, and overlay jet contours at the corresponding times from the simulations in which jets were launched. For $\mathfrak{M} = 4$ (top panels), the jet morphology closely follows the underlying ISM structure. This indicates, first, that jet propagation is strongly controlled by the highly turbulent ambient medium and, second, that the ISM can continuously resupply gas to the black hole, highlighting the importance of ISM-driven regulation. For the lower Mach number case, $\mathfrak{M} = 1$ (bottom panels of Fig.~\ref{fig:machcontours}), the jet sweeps away the material that would otherwise reside near the black hole. This suggests that in less turbulent media the jet more strongly modifies the gas distribution and, as a result, more effectively self-limits the accretion rate.

These results indicate that turbulence plays \textit{a dual role}: it facilitates both the outward propagation of jet material and the inward delivery of gas that fuels accretion and powers the jet.

\subsection{Dependence on the opening angle}

AGN outflows are not always collimated, particularly at low-luminosity \citep{Elvis2000}.
Therefore, we investigate how the opening angle of an AGN outflow affects accretion. To bracket plausible outflow geometries, we consider two limiting cases: a collimated jet and a spherically symmetric wind, i.e. $\theta = 0^\circ$ and $\theta = 180^\circ$, respectively.

\begin{figure}[ht!]
    \centering
    \includegraphics[width=\linewidth]{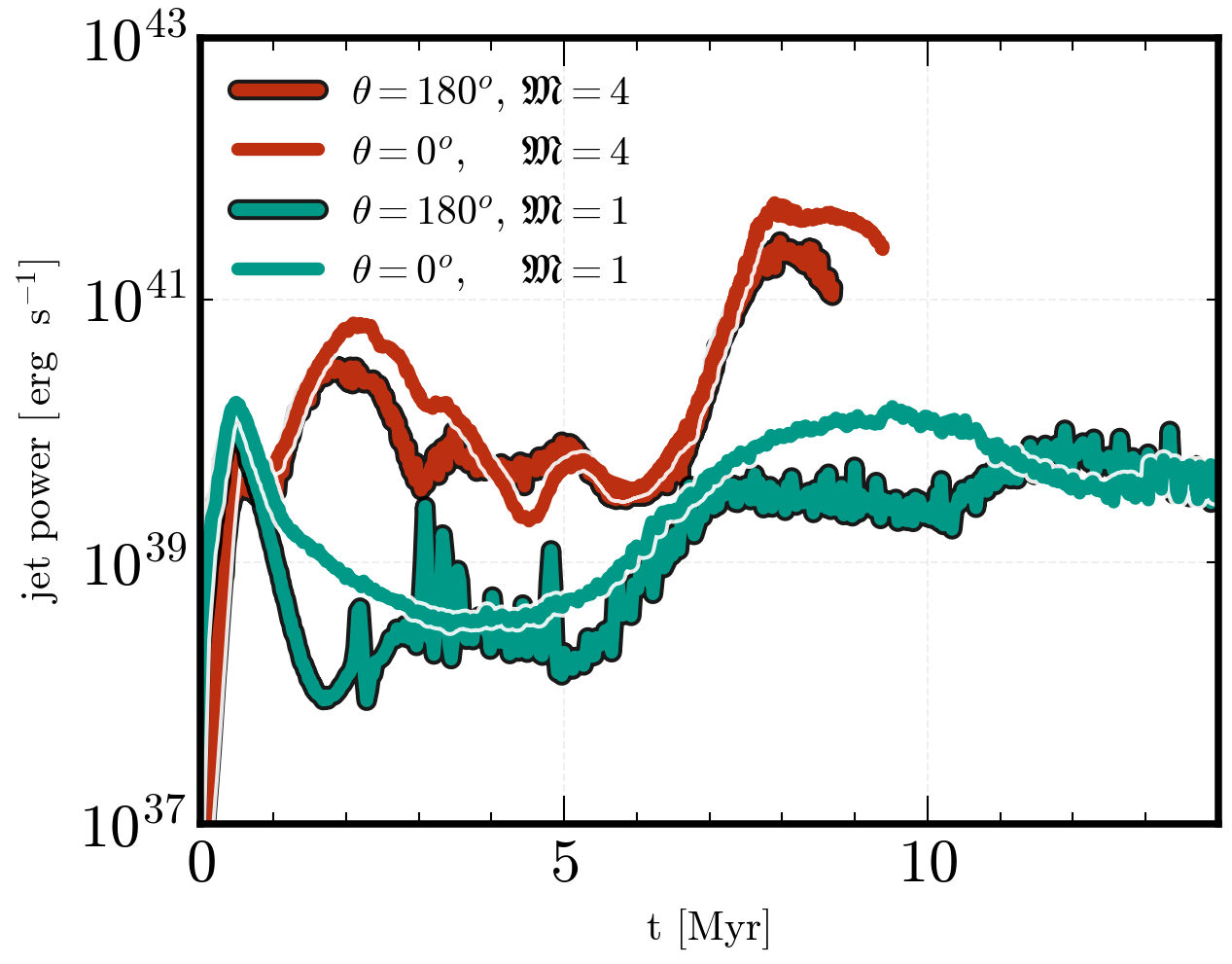}
    \caption{Jet power evolution for models with different jet opening angles. All simulations have the same ISM number density $n = 20$ cm$^{-3}$ and jet efficiency $\varepsilon=10^{-2}$, but we show results for two different Mach numbers, $\mathfrak{M} = [1, 4]$.}
    \label{fig:angle}
\end{figure}

In Fig.~\ref{fig:angle}, we show jet power evolution for simulations with AGN modeled as collimated jets or spherical winds. 
We observe that increasing the opening angle reduces both the black hole accretion rate and the injected jet power; for $\mathfrak{M}=1$ and $\mathfrak{M}=4$, the average jet power decreases by $\sim30\%$ and $\sim15\%$, respectively. 
A spherical wind, which pushes gas outward in all directions, more effectively suppresses inflow, whereas a collimated jet can still be fed by gas accreting from directions transverse to the jet axis. 
Thus, a wind can limit accretion by reducing the inflow, but it does not fully quench it in our runs. For example, for $\mathfrak{M}=1$, between 1 and 2\,Myr, the spherical wind shows a pronounced decline in power consistent with self-limiting accretion, while the collimated jet remains more steadily fueled. Overall, larger opening angles lead to lower accretion rates and promote self-limiting accretion.

\subsection{Connection to FR0 sources} \label{FR0}

In the previous sections, we showed that the dynamic pressure of a turbulent medium can stall the jet on small scales. Moreover, jets launched in turbulent medium develop asymmetric and complex morphologies (see, e.g., Fig.~\ref{fig:machcontours}), similar to those observed in low-luminosity and stalled AGN \citep{Giovannini2023}. 
In addition to the morphological resemblance to FR0 sources, we can also compare the quantitative properties of the jets.

\begin{figure}[ht!]
    \centering
    \includegraphics[width=\linewidth]{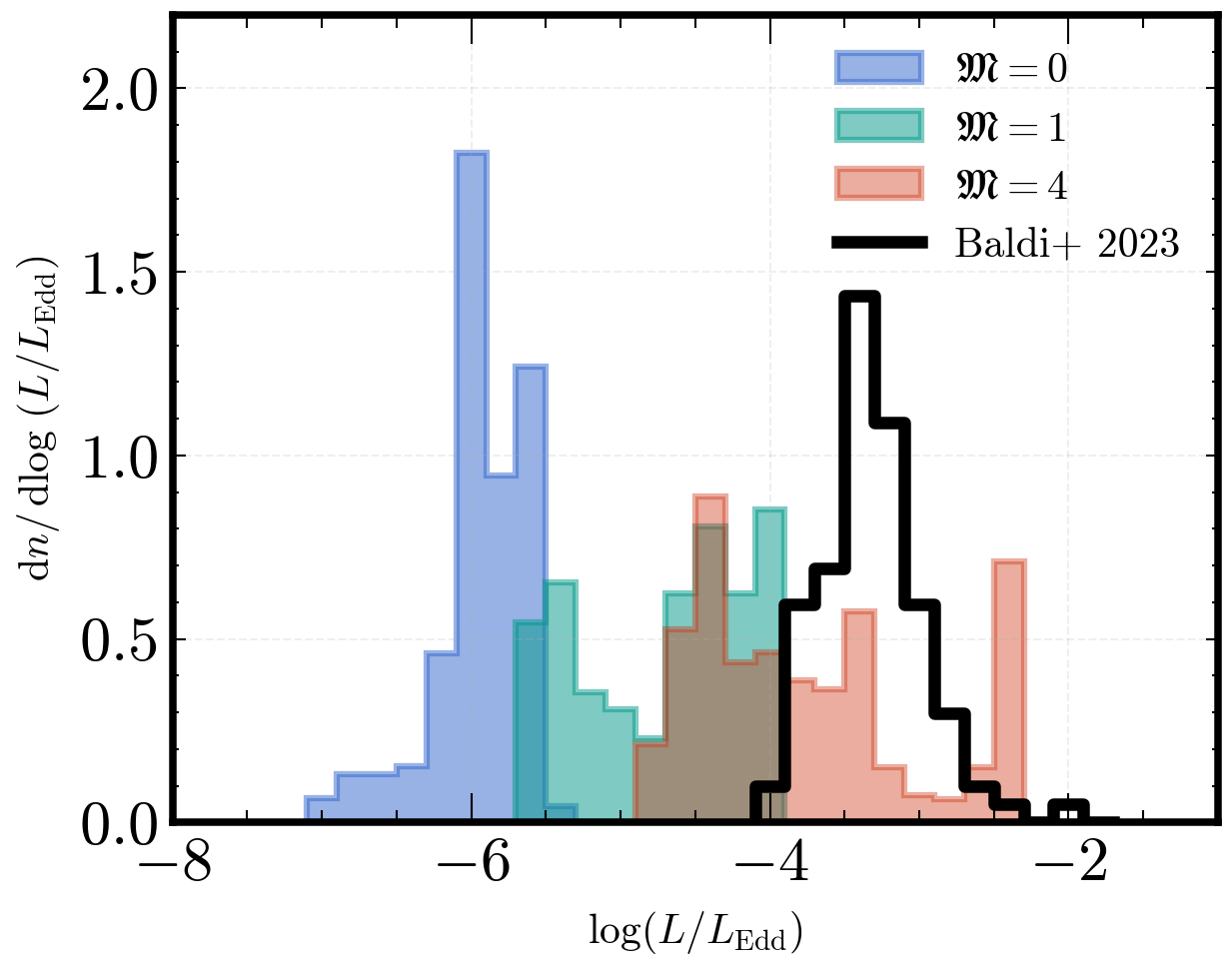}
    \caption{
Distribution of Eddington ratios. The solid line shows the observed distribution of $(L_{\mathrm{bol}} + L_{\mathrm{mech}})/L_{\mathrm{Edd}}$ from \citet{Baldi2023}. The filled histograms correspond to Eddington ratio $L_{\rm jet}/L_{\rm Edd}$ for the jets launched in media with $\mathfrak{M} = 0, 1, 4$ and number density $n = 20\,\mathrm{cm^{-3}}$, with jet parameters $\varepsilon = 10^{-2}$ and opening angle $\theta = 0^\circ$.
    }
    \label{fig:fr0}
\end{figure}

In Fig.~\ref{fig:fr0}, we compare the distribution of Eddington ratios for FR0 sources at low redshift $z<0.05$ \citep{Baldi2018} with those obtained from our simulations. \citet{Baldi2023} estimated the bolometric luminosity $L_{\mathrm{bol}}$ and mechanical jet power $L_{\mathrm{mech}}$ from radio observations and computed the Eddington ratio as $\frac{L_{\mathrm{bol}} + L_{\mathrm{mech}}}{L_{\mathrm{Edd}}}$. In our analysis, we calculate the distribution of jet power in simulations with $\mathfrak{M} = 0, 1, 4$ by sampling the time evolution of $L_{\mathrm{jet}}$ across snapshots, and normalize it by the Eddington luminosity $L_{\mathrm{Edd}}$ corresponding to a black hole mass of $10^{6}\,M_\odot$.

We find that the distribution of Eddington ratios for FR0 sources most closely approaches the distribution obtained for jets propagating in a medium with $\mathfrak{M} = 4$. The largest discrepancy is found for jets launched in a uniform medium. We performed tests with higher efficiency factors for jets launched in uniform media, and we found that, even then, due to self-limiting accretion, Eddington ratios remain in the same range.

We note, however, that the FR0 catalog includes black holes with masses above $10^{7}\,M_\odot$, whereas our simulations assume a black hole mass of $10^{6}\,M_\odot$. In our model, the efficiency factor is degenerate with black hole mass, as follows from the BHL formula (Eq.~\ref{eq:Bondi}): increasing the black hole mass by a factor of 10 is equivalent to increasing the efficiency factor by 100. So, on the one hand, in a non-turbulent case with Mach number $\mathfrak{M}=0$, where the accretion is self-limited, the jet power is independent of efficiency factor or, as we note above, of the black hole mass.
On the other hand, $L_\mathrm{Edd}$ scales linearly with black hole mass, so increasing the mass by a factor of 10 raises $L_\mathrm{Edd}$ by the same factor. As a result, for $\mathfrak{M}=0$, the ratio $L/L_\mathrm{Edd}$ decreases by a factor of 10, shifting the Eddington-ratio distribution toward lower values and further increasing the discrepancy with the FR0 sample.

In a turbulent medium, by contrast, the accretion rate still responds to the black hole mass. For $\mathfrak{M}=4$, increasing the mass by a factor of 10 raises the jet power by a factor of 100 (Eq.~\ref{eq:Bondi}) and hence the Eddington ratio by a factor of 10, reaching a value of $f_\mathrm{Edd} \sim 10^{-1.5}$. A sufficiently large mass would, however, eventually drive the system into the self-limiting regime, beyond which the Eddington ratio would no longer increase.

We stress that we cannot fully reproduce the FR0 population due to the limitations of our model. We do not model the galactic potential, nor do we include a full galactic environment, which means that inflows to the black hole are driven by turbulence alone rather than by gravitational pull. This may be especially important for jet evolution on timescales longer than $\sim13$\,Myr.

To conclude, our uniform-medium simulations do not approach the Eddington ratios observed in FR0 sources, whereas jets launched in a turbulent, multiphase medium approach them more closely. A fair comparison, however, would need to account for both the observational constraints and the idealized nature of our simulations.

\section{Discussion} \label{sec:discussion}

In this section, we place our results in the context of previous work. We first compare our simulations with earlier numerical studies that implement self-consistent jet launching, and we then relate our findings to observational constraints. Finally, we discuss the limitations of our model and their implications.

\subsection{Comparison with simulations}

In our previous work with fixed jet power \citep{Borodina2025}, we showed that the ISM density strongly affects jet propagation: denser gas produces higher dynamical pressure and can stall the jet more effectively. Once we allow the jet power to respond to the ambient medium, however, we find that $L_{\mathrm{jet}}$ scales approximately linearly with the ISM density. This compensates for the increased confining pressure and renders jet propagation largely insensitive to the mean density.

\citet{Talbot2021, Talbot2022} model self-consistent jet launching with the jet power computed from a Blandford–Znajek-based prescription \citep{Blandford1977}, rather than the simple Bondi-like scaling adopted in our work, and study the jet's interaction with a resolved circumnuclear disc rather than a larger-scale ISM. \citet{Talbot2021} showed that self-consistently launched jets develop morphologies that differ from those in fixed–jet-power simulations. In our study, for efficiency factors below the self-limiting accretion regime (e.g. $\varepsilon \lesssim 10^{-1}$ for an ISM with Mach number $\mathfrak{M} = 4$ and $\varepsilon \lesssim 10^{-3}$ for $\mathfrak{M} = 1$ ), we instead recover morphologies similar to our fixed-power jets \citep{Borodina2025}, because in this regime the ISM primarily regulates jet propagation. \citet{Talbot2022} further demonstrated that the environment surrounding the black hole can significantly alter jet morphology, a trend we also find in our simulations. Despite differences in setup and included physics, \citet{Talbot2022} report a broadly similar range of jet powers, $10^{40}$–$10^{43}\,\mathrm{erg\,s^{-1}}$.

Moreover, \citet{Cammelli2026a, Cammelli2026b} study jet-regulated accretion, launching a self-consistent kinetic jet in a turbulent medium, as in our work. However, their setup differs from ours in the included physics: it incorporates a gravitational potential and radiative cooling with a cold-phase prescription, neither of which we include, and spans a larger dynamic range (kpc to sub-pc) than ours. \citet{Cammelli2026a, Cammelli2026b} also study the maintenance (radio) mode in a quenched galaxy-group atmosphere, which is a hot, diffuse medium, rather than the dense, gas-rich turbulent ISM we consider. As a result, their jets remain collimated as they propagate through the halo, whereas ours interact with, and are frequently confined by, the surrounding dense ISM. The two setups thus probe complementary feeding–feedback regimes. Despite these differences, we recover a common trend: stronger turbulence drives greater variability in the accretion rate and jet power.

A more subtle comparison is with cosmological simulations, where most of the ISM turbulence is ultimately driven by stellar feedback. We do \emph{not} model stellar feedback explicitly; instead, we impose turbulence at high resolution in a closed box with a fixed gas supply. As a result, our simulations capture small-scale turbulent structure but do not include large-scale gas removal by winds or outflows. In contrast, cosmological zoom-in simulations with typical mass resolutions of $\sim 3\times10^{4}\,M_\odot$ \citep[e.g.,][]{AnglesAlcazar2017, Byrne2023} cannot fully resolve the turbulent cascade down to Bondi scales. In those models, stellar feedback primarily manifests as bulk removal of gas from the central regions rather than as a well-resolved, turbulent ISM. 
As a result, they show that stellar feedback leads to accretion regulation.
They also omit SMBH feedback, so accretion is limited by stellar feedback–driven outflows alone.

Within our setup, we find that turbulence \emph{boosts} accretion: a more turbulent environment---such as one that might be produced by strong stellar feedback---more efficiently delivers dense gas to the accretion region and can enhance jet power. In the presence of AGN feedback, the jet can temporarily shut off accretion onto the black hole, while turbulence subsequently resupplies material to the injection region.

Cosmological simulations that include both stellar and AGN feedback \citep{Dubois2015, Habouzit2017} find that supernova (SN) feedback plays a crucial role in suppressing black hole growth in low-mass galaxies with shallow gravitational potentials, primarily by ejecting gas from the central regions.
\citet{Koudmani2022} found, however, that the strength of this regulation is partly a modelling choice: strong supernova feedback has historically been tuned to reproduce the low star-formation efficiency of low-mass galaxies. These authors showed that when this assumption is relaxed, moderate supernova feedback combined with efficient AGN feedback can regulate star formation in dwarf galaxies, suggesting that AGN may play a larger role than is usually assumed.
For more massive systems, by contrast, there is broad consensus that AGN feedback dominates. Once a galaxy reaches a stellar mass of $M_* \sim 10^9\,M_\odot$, corresponding to a black hole mass of $M_{\rm BH} \sim 10^6\,M_\odot$, black hole growth becomes predominantly regulated by AGN feedback \citep{Dubois2015, Habouzit2017}. Since our simulations adopt $M_{\rm BH}=10^6\,M_\odot$ and do not allow gas to escape the box, our results are more relevant to this latter, AGN-dominated regime and do not contradict the view that SN feedback limits accretion at earlier stages when gas removal is efficient.

Moreover, cosmological zoom-in simulations that include AGN feedback, such as \citet{Bonoli2016}, find that for systems hosting SMBHs with masses $\sim 10^6\,M_\odot$, feedback remains confined to the central regions and can lead to black hole starvation for a total efficiency factor of $\varepsilon = 5\times 10^{-3}$.  In our simulations, for $\mathfrak{M} = 4$ we only observe clear self-limiting accretion at higher efficiencies, $\varepsilon > 10^{-1}$, at which point the jet is no longer confined within the central kiloparsec. At lower Mach numbers, however, both confinement and self-limiting accretion can occur already for $\varepsilon \sim 10^{-3}$, which is consistent with the findings of \citet{Bonoli2016}.

As a result, our closed-box setup, in which feedback cannot drive large-scale gas loss, is therefore best interpreted as probing how AGN jets interact with an already gas-rich central ISM, rather than as a full model for the long-term, feedback-regulated co-evolution of galaxies and their central black holes.

\subsection{Comparison with observations}

We recover a range of jet powers consistent with those observed in compact jet systems \citep{Roy2021, Fabbiano2022}. For the Eddington ratios, our aim is not to reproduce the FR0 distribution exactly but to identify the conditions under which it can be approached: only our supersonic-turbulence runs reach Eddington ratios within an order of magnitude of those reported in the FR0 catalog \citep{Baldi2018, Baldi2023}, whereas jets launched into a uniform, non-turbulent medium have systematically lower power. This suggests that at least mildly supersonic turbulence is required to reach the observed FR0 regime. However, several caveats apply to this comparison. The black hole mass adopted in our simulations is lower than that of the catalog sample, but a higher mass would not bring the distribution closer to the observed values as we discussed in Section~\ref{FR0}. 
In addition, the FR0 catalog is flux-limited and therefore biased toward higher Eddington ratios, so the observed distribution is not representative of the whole population of FR0s. We also note that the jet power inferred from observations are not directly comparable to the jet powers measured in our simulations.
Moreover, our setup does not include a host-galaxy potential, so the gas reservoir near the Bondi radius is sustained by the periodic boundaries and the imposed turbulent driving rather than by gravitationally driven inflow; the Eddington ratios we recover therefore identify the conditions under which the FR0 regime can be approached, not a gravitationally regulated steady state. A more rigorous comparison would therefore involve both constructing mock radio images and applying the same observational calibrations, and modeling a full galactic context, all of which is beyond the scope of this work.

We also show that jets with efficiency factors $\varepsilon < 10^{-1}$ remain confined within the central $\sim$1\,kpc for several Myr while maintaining jet powers above $10^{38}\,\mathrm{erg\,s^{-1}}$, consistent with the compactness and low-to-moderate mechanical powers inferred for FR0 sources \citep{Baldi2015, Baldi2016, Baldi2023}. In this regime, the jet interacts with the dense, multiphase ISM and inflates small cavities without rapidly escaping the host, allowing sustained energy injection into the surrounding gas though it remains unclear whether this is sufficient to significantly affect the host galaxy on kiloparsec scales (see e.g. \citet{Borodina2025}). At higher efficiencies, the jets propagate beyond the central kiloparsec and can strongly disrupt the ISM, potentially driving larger-scale outflows similar to those observed in red geyser systems \citep{Roy2021}.

We further show that turbulence naturally produces asymmetric and bent jet structures. In our models, turbulent density fluctuations create pre-existing low-density channels and cavities that guide the jet, leading to intermittent propagation and strongly asymmetric morphologies even at fixed jet parameters. This behavior provides a natural explanation for the complex, one-sided, or distorted structures observed in compact, low-power radio jets \citep{Fabbiano2022, Giovannini2023}, and supports the idea that environmental inhomogeneity is a key ingredient for reproducing FR0-like morphologies and energetics.

\subsection{Limitations of our model}

Our simulations omit several physical processes that could influence both jet propagation and black hole accretion. We briefly summarize the most important limitations here.

First, we do not include magnetic fields or cosmic rays. Both can alter the interaction between the jet and the ambient medium, for example by affecting mixing, transport, and the effective pressure support of the gas, and may therefore modify the morphology and efficiency of feedback. These non-thermal components will be incorporated in future work.

Second, our ISM model lacks an explicitly cold molecular phase. Allowing gas to cool below $10^4$\,K would likely make the medium less filamentary and more clumpy, which could facilitate jet propagation through low-density channels while also providing dense reservoirs that may enhance accretion onto the black hole. The net impact on the coupling between the jet and the ISM remains uncertain and warrants dedicated investigation. Nevertheless, resolving a multiphase medium down to molecular cloud scales in the presence of a self-consistently launched jet is computationally expensive, and is beyond the scope of this study.

Third, we do not include the effects of gravity.
In a realistic system the host galaxy's gravitational potential plays two roles that our periodic box cannot capture: it drives the large-scale inflow that replenishes gas at the Bondi radius, and it provides the restoring force that returns displaced or cooled gas toward the black hole after a feedback episode. In our simulations the gas density near the Bondi radius is instead maintained by the periodic boundaries and the imposed turbulent driving, so the resupply of the accretion region is not gravitationally driven.
The timescale on which this omission matters can be estimated from the free-fall time: for $10^4$\,K gas and a $10^6\,M_\odot$ black hole, it is $\sim$13\,Myr at the Bondi radius ($\sim$90\,pc), comparable to the duration of our runs, and exceeds 100\,Myr near the edge of the box.
Introducing a realistic galactic potential in similar setups is nontrivial,  as the system can collapse on relatively short timescales once cooling and gravity are both active. Nonetheless, gravity will influence gas inflows, disc structure, and the confinement of jet-inflated cavities, and thus can affect both the accretion rate and the resulting feedback.

Finally, our accretion prescription does not explicitly depend on the black hole spin or on the angular momentum of the gas reaching the black hole. Spin can enter implicitly through the choice of efficiency factor, but we do not model its evolution or orientation. This is justified because the spin evolves on a timescale much longer than that of our simulations: applying Eq.~17 of \citet{Narayan2022}, we find that even for our highest-Eddington-ratio run the spin changes by less than 0.1\% over the $10$\,Myr duration of our simulations. We also do not track the detailed angular momentum distribution of the inflowing gas that would determine disc structure and jet launching in more realistic setups. Including these effects, together with gravity and a cold phase, could increase the accretion rate in some regimes and alter the time variability and power of the jets. These extensions are important directions for future work.

\section{Conclusion} \label{sec:conclusion}

In this paper, we present idealized simulations of AGN jets propagating through a turbulent ISM. We investigate how the properties of the ISM and jet-launching parameters influence jet propagation and the coupled evolution of accretion and feedback. We characterize the ISM using the mean number density and turbulent Mach number, and the jet using the efficiency factor and opening angle. By varying these four parameters, we explore the relative roles of the ISM and black holes in shaping the regulation and evolution of jets.

We find that jet propagation is largely insensitive to the average background density. While a denser environment leads to higher accretion rates and stronger jets, the turbulent ISM structure controls jet propagation. As a result, the higher jet power in denser media primarily increases the total injected energy and heats the gas inside existing cavities more efficiently, but it does not substantially extend the cavities to larger radii.

The ISM Mach number strongly influences both the evolution of jet power and the jet’s ability to propagate. For Mach numbers $\mathfrak{M} \leq 2$, variations in jet power increase as turbulence becomes stronger. However, at $\mathfrak{M} = 4$, this trend saturates: the variability does not continue to grow. We attribute this behavior to rapid gas replenishment in the vicinity of the black hole caused by the highly dynamic medium.

Supersonic turbulence generates an ISM with a broad density distribution, in which dense filaments and walls lead to strong boosts in jet power, while low-density cavities allow the jet to propagate to larger distances by filling them. In contrast, subsonic turbulence and a uniform medium confine the jet more effectively, even though the jet power remains relatively low and stable. In our closed-box, high-resolution setup --- where we impose turbulence rather than generating it self-consistently through stellar feedback --- this turbulence \emph{enhances} accretion and jet power. This differs from cosmological simulations, in which limited resolution and efficient stellar-feedback–driven gas removal tend to suppress accretion.

Jet-launching parameters also play a significant role in jet evolution. Whether the jet reaches a self-limiting accretion regime depends on the efficiency factor $\varepsilon$, and the threshold shifts with the turbulent Mach number. For the $\mathfrak{M} = 4$ ISM, at low efficiencies ($\varepsilon \lesssim 10^{-1}$) we see no clear self-limiting behavior, and temporal variations in jet power are instead set by the state of the surrounding medium; at higher efficiencies ($\varepsilon > 10^{-1}$), the jet can enter a self-limiting regime by clearing the gas that would otherwise fuel the SMBH, though the most powerful jets do not remain confined within the central kiloparsec over the $\sim 5\,$Myr we simulate. At lower Mach number this transition shifts to smaller efficiencies: for $\mathfrak{M} = 1$, the jet power converges by $\varepsilon \simeq 10^{-3}$, while the jets remain confined within the box throughout our simulations (see Appendix~\ref{app:self-lim-M}).

The opening angle of the outflow affects the accessibility of gas to the black hole and systematically reduces the accretion rate as the outflow becomes less collimated, but does not fully suppress inflow in either turbulent or uniform media.

Finally, turbulent environments appear to approach the properties of FR0 sources more closely than a uniform medium does. In particular, we find that the asymmetric jet morphologies characteristic of these systems are difficult to reproduce in a uniform medium, suggesting that turbulence in the ISM plays an important role. In addition, jets launched in a uniform environment tend to lie at much lower Eddington ratios than those typically observed in FR0 sources. This mismatch is not resolved by increasing the jet efficiency (because of self-limiting accretion) or by increasing the SMBH mass (because the Eddington luminosity increases), pointing to at least mildly supersonic turbulence as a necessary condition for approaching the observed Eddington-ratio regime. We emphasize that, in the absence of a host-galaxy gravitational potential, this comparison remains qualitative: a definitive match to the FR0 population would require modeling the full galactic context alongside mock radio observations.

\begin{acknowledgments}
This work was supported by the Simons Collaboration on “Learning the Universe”.
We thank the anonymous referee for their constructive comments. We thank Charlie Conroy and Lisa Kewley for their insights into the physical role of AGN feedback, and Samuel Ward, Hyerin Cho, Matthew Smith, and Rosie Talbot for their comments on the simulation aspects of this work. We also thank Alexander Marchuk for the paper title inspiration.

The authors used Claude (Anthropic) to assist with grammar and language editing. All scientific content and conclusions are the authors’ own, and the authors take full responsibility for the final text.

\end{acknowledgments}

%



\software{Arepo \citep{Springel2010},
          Gaepsi \citep{Gaepsi2011}}


\appendix

\section{Density and temperature distributions for different Mach numbers} \label{app:hists}

In this appendix, we present how the temperature and number density distributions depend on the Mach number. For each simulation with the same mean number density, $n = 20\,\mathrm{cm^{-3}}$, and Mach numbers $\mathfrak{M} = [0.5, 1, 2, 4]$, we select snapshots taken before the jet is launched, at the times when the turbulent boxes have reached a steady state (see Table~\ref{tab:turbbox}). We then compute the mass-weighted distributions of temperature and density. 

\begin{figure}[ht!]
    \centering
    \includegraphics[width=\linewidth]{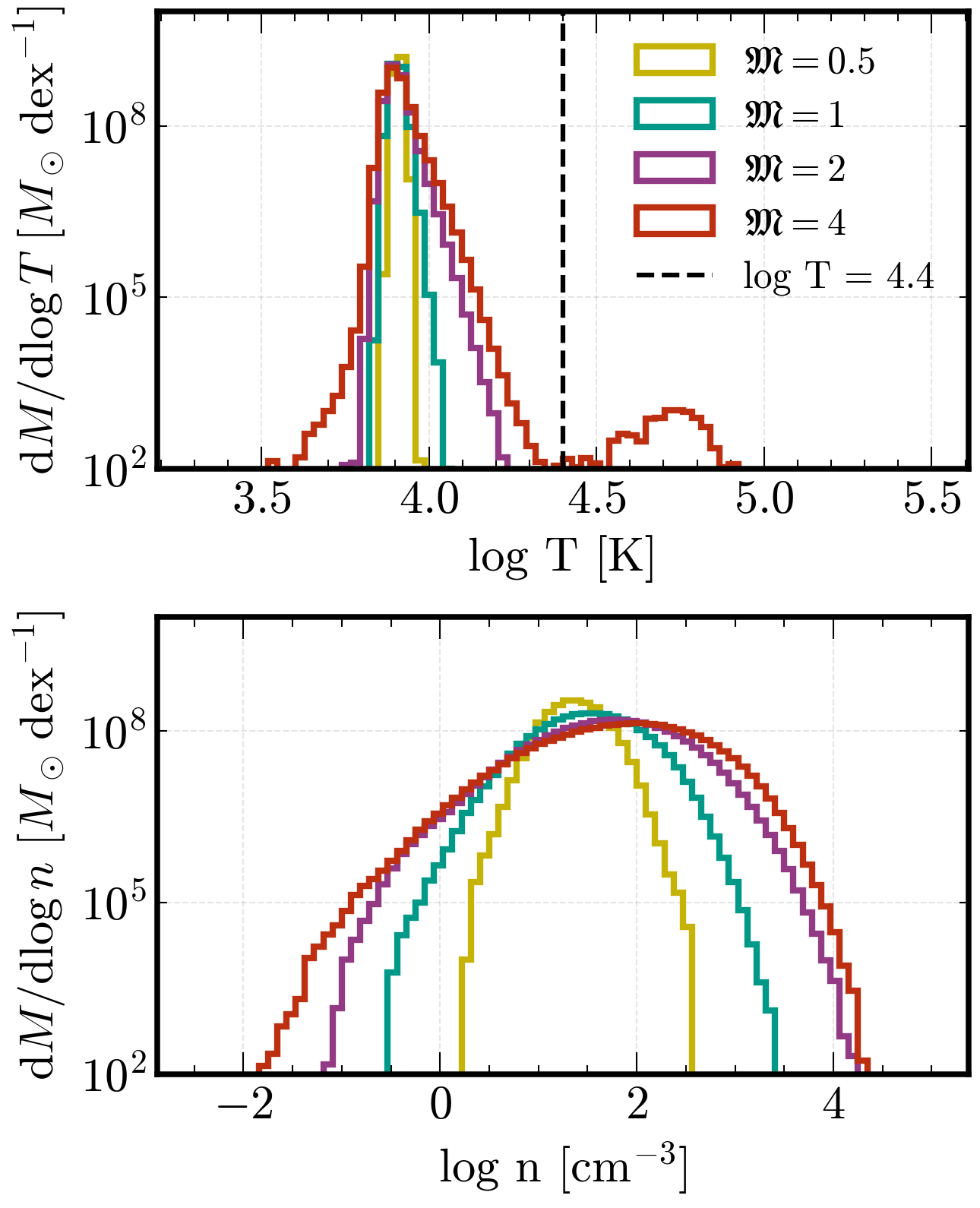}
    \caption{
    Top: Mass-weighted temperature distributions of gas cells for simulations with different Mach numbers, as indicated by color. The black dashed line marks the temperature threshold used to distinguish between gas phases. Bottom: Mass-weighted number density distributions for the same snapshots as in the top panel.
    }
    \label{fig:hists}
\end{figure}

As shown in Figure~\ref{fig:hists}, as the turbulence strength increases, the distributions of both temperature and density broaden. For $\mathfrak{M} = 4$, the gas begins to develop a distinct phase with $\log T > 4.4$.

\section{Self-limiting accretion for lower mach number case} \label{app:self-lim-M}

\begin{figure*}[ht!]
    \centering
    \includegraphics[width=\textwidth]{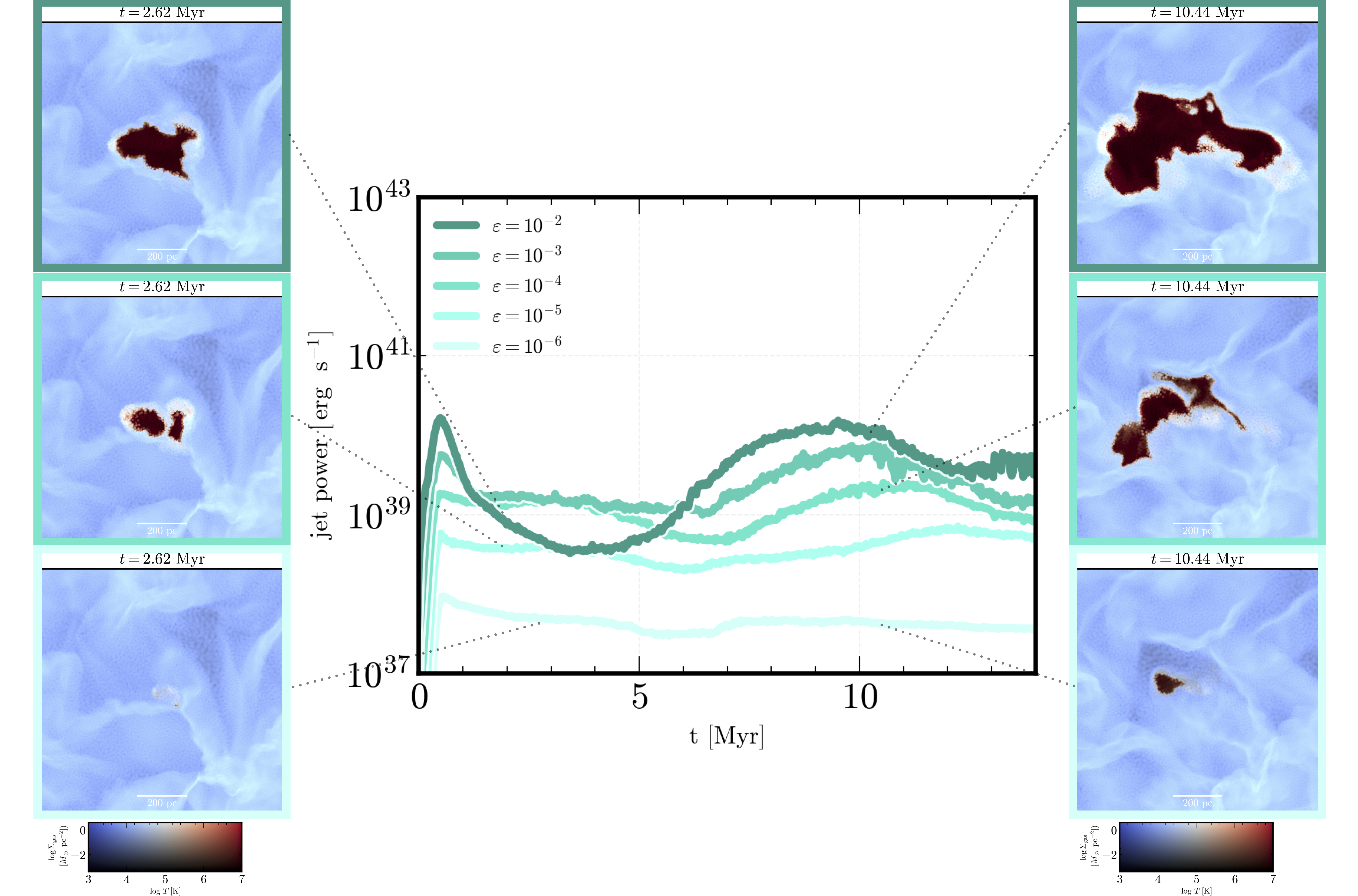}
    \caption{\textit{Central panel}: Jet power evolution for models with different jet efficiency factor values.
    \textit{Left panels}: Density-temperature projection maps of simulations with $\varepsilon=[10^{-2}, 10^{-4}, 10^{-6}]$ at 2 Myr after the start of jet launching.
    \textit{Right panels}: Density-temperature projection maps of simulations with $\varepsilon=[10^{-2}, 10^{-4}, 10^{-6}]$ at 10 Myr after the start of jet launching.    
    All the simulations have the same ISM conditions: $n = 20$ cm$^{-3}$, $\mathfrak{M} = 1$. The jet opening angle is set to $0^\circ$.}
    \label{fig:epsilon_m1}
\end{figure*}

In this appendix, we show that jet's ability to limit accretion depends on the efficiency factor $\varepsilon$ also for simulations with Mach number $\mathfrak{M} = 1$. In Section~\ref{sec:results_epsilon}, we demonstrated this effect at higher Mach number. However, in those simulations the most powerful jet reaches the box boundary within the first 4\,Myr, forcing us to stop the simulation shortly after the convergence of jet power with $\varepsilon$ starts to appear. To show this convergence more clearly, we performed additional simulations at lower Mach number (see Fig.\,\ref{fig:epsilon_m1}). For $\mathfrak{M} = 1$, jet is not powerful enough to reach the box boundary within the 14\,Myr of our simulations (see projection maps in Fig.\,\ref{fig:epsilon_m1}), precisely because accretion becomes self-limited at lower efficiency factors.

As shown in Fig.\,\ref{fig:epsilon_m1}, for $\mathfrak{M} = 1$ and  $\varepsilon<10^{-4}$, the jet power scales almost linearly with the efficiency factor. Convergence sets in at $\varepsilon=10^{-3}$, and at the higher value of efficiency factor, $\varepsilon=10^{-2}$, the jet power can even drop below that of lower-efficiency runs (visible between 1 and 6\,Myr), as the more powerful jet more strongly suppresses its own fuel supply.

\section{Sensitivity to turbulent seed} \label{app:seeds}

\begin{figure*}[ht!]
    \centering
    \includegraphics[width=0.9\textwidth]{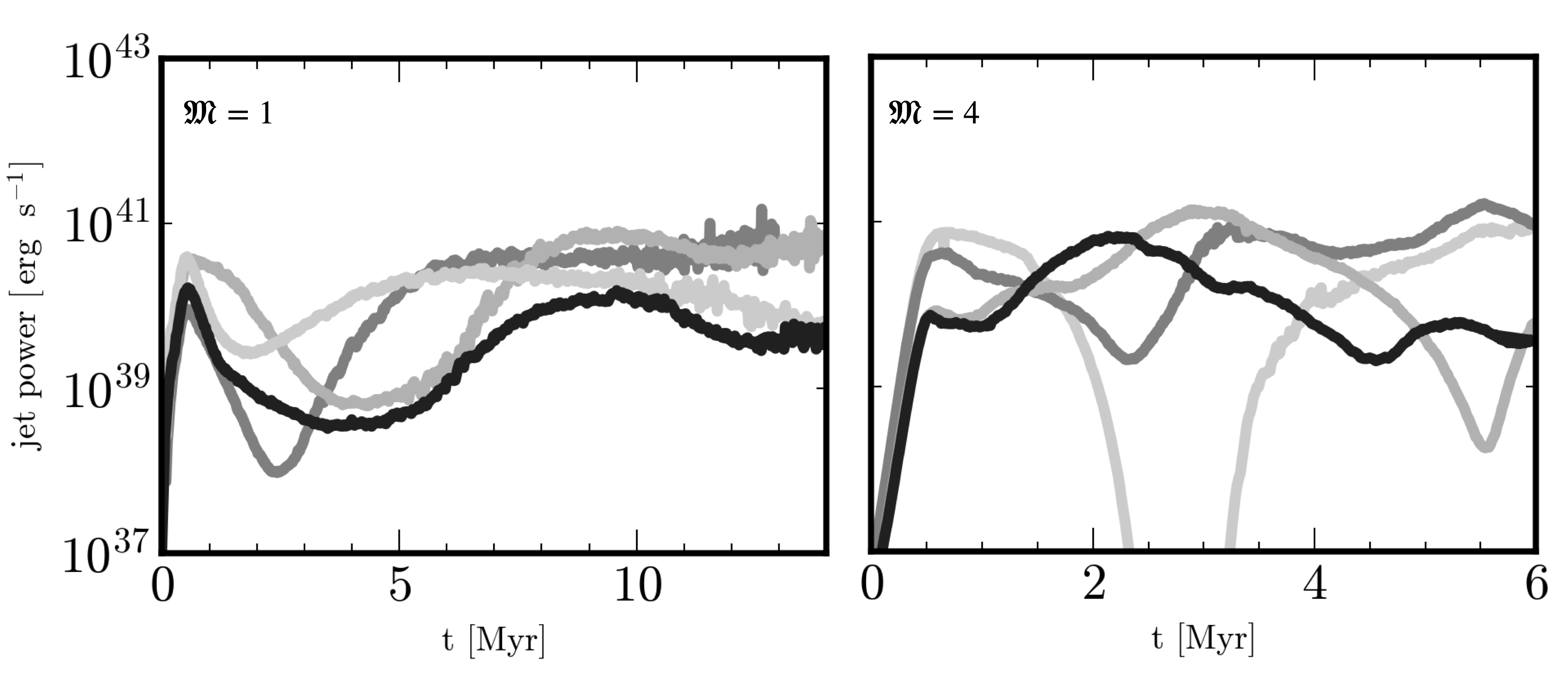}
    \caption{Jet power evolution for models with different turbulent seeds. 
    \textit{Left panel:} The jets are launched into a medium with Mach number $\mathfrak{M} = 1$ and $n = 20\,\mathrm{cm^{-3}}$. Jet has efficiency of $\varepsilon = 10^{-2}$ and opening angle is set to $\theta = 0^\circ$.
    \textit{Right panel:} The jets are launched into a medium with Mach number $\mathfrak{M} = 4$ and $n = 20\,\mathrm{cm^{-3}}$. Jet has efficiency of $\varepsilon = 10^{-2}$ and opening angle is set to $\theta = 0^\circ$.
    }
    \label{fig:seeds}
\end{figure*}

To verify that our results do not depend on the particular realization of the turbulent velocity field, we repeated a subset of our simulations with different random seeds, holding all other parameters fixed. Figure~\ref{fig:seeds} shows the jet-power evolution for four seeds in each case, at $\mathfrak{M} = 1$ (left) and $\mathfrak{M} = 4$ (right).

For both Mach numbers, the instantaneous jet power depends on the seed, since it is governed by the stochastic passage of turbulent density fluctuations through the accretion region. This dependence is strongest in the supersonic case ($\mathfrak{M} = 4$, right panel), where individual realizations show large temporal variations. Despite these fluctuations, all seeds share the same characteristic power level
($\sim 10^{39}$--$10^{41}\,\mathrm{erg\,s^{-1}}$) and time-averaged behavior.

At $\mathfrak{M} = 1$ (left panel), the jet power evolution is smoother. Unlike the supersonic case, after an initial peak at $\sim 10^{40}\,\mathrm{erg\,s^{-1}}$ each realization undergoes a pronounced dip as the jet suppresses its own accretion supply. Although the depth and length of this dip vary from seed to seed, all realizations converge to a similar quasi-steady jet power of order $10^{40}\,\mathrm{erg\,s^{-1}}$ by $\sim 10\,\mathrm{Myr}$, roughly an order of magnitude below the supersonic case.

In both cases the turbulent seed affects only the detailed timing of individual fluctuations, not the quantities. As a result, our conclusions are robust to the choice of turbulent realization.




\bibliography{jet-ism}{}
\bibliographystyle{aasjournal}



\end{document}